\documentstyle[11pt,newpasp,twoside]{article}
\def\edcomment#1{\iffalse\marginpar{\raggedright\sl#1\/}\else\relax\fi}
\reversemarginpar

\begin{document}

\title{Oscillations of rapidly rotating stars}
\author{B. Dintrans \& M. Rieutord}
\affil{Observatoire Midi-Pyr\'{e}n\'{e}es, 14 avenue Edouard Belin,
F-31400 Toulouse. E-mail: dintrans@obs-mip.fr}

\begin{abstract}
We present numerical simulations of gravito-inertial waves propagating in 
radiative zones of rapidly rotating stars. A first model, using the 
Boussinesq approximation, allows us to study the oscillations of a 
quasi-incompressible stratified fluid embedded in a rapidly rotating sphere 
or spherical shell. In a second step, we investigate the case of a $\gamma$ 
Doradus-type star using the anelastic approximation. Some fascinating 
features of rapidly rotating fluids, such as wave attractors, appear 
in both cases.
\end{abstract}

\section{The Boussinesq model}

In this configuration, the Brunt-V\"ais\"al\"a\ frequency is simply
proportional to the radial distance. Many mathematical results are known
(Friedlander \& Siegmann 1982) concerning the shape of critical surfaces
and characteristics of the governing mixed type operator. We confirm all 
these results by calculating the orbits of characteristics which propagate 
in the hyperbolic domain. In particular, we find that characteristics can be
focused along attractors leading to associated singular velocity fields
(Dintrans et al. 1999).

\section{The anelastic model}

We study, using the anelastic approximation, the low-frequency 
oscillations of a typical $\gamma$-Doradus star. Hence, we show that 
dealing with rotation by the means of a second-order perturbative theory 
is not correct for rotation periods less than 3 days. Using the same 
geometric formalism as above (i.e. calculations of orbits of characteristics) 
allows us to compute the frequencies of oscillations in the rapid rotation 
r\'egime (rotation periods $\sim$ 1 day). We find again that wave 
attractors are a common feature of rapidly rotating fluids and conclude that
they are promising features for the transport of angular momentum and
chemicals in the radiative zone of rotating stars (Dintrans \& 
Rieutord, 1999).


\begin{references}
\reference Dintrans B., Rieutord M., Valdettaro L., 1999, J. Fluid Mech., 
in press
\reference Dintrans B., Rieutord M., 1999, \aap, in press
\reference Friedlander S., Siegmann W. L., 1982, J. Fluid Mech., 114, 123
\end{references}
\end{document}